\documentclass[a4paper]{spie}

\usepackage{graphicx}
\usepackage{textcomp}

\title{The development of the THESEUS SXI optics}

\author[a]{Charlotte Feldman}\author[a]{Paul O'Brien}\author[a]{Richard Willingale}
\author[b]{Emile Schyns}\author[b]{Romain Roudot}\author[b]{Ray Fairbend}\author[b]{Julien S\'{e}guy}
\author[a]{Hannah Lerman}\author[a]{Ian Hutchinson}\author[a]{Melissa McHugh}\author[a]{Alexander Lodge}\author[a]{Roisin Speight}
\affil[a]{University of Leicester, University Road, Leicester, LE1 7RH, UK}
\affil[b]{Photonis France S.A.S., Avenue Roger Roncier, 19100 Brive, B.P. 520, 19106 Brive Cedex, France}

\authorinfo{Further author information: Charly Feldman: \\E-mail: chf7@le.ac.uk, Telephone: +44 (0)116 252 5084}
 
\begin{document} 
\maketitle

\begin{abstract}
The Transient High Energy Sources and Early Universe Surveyor is an ESA M5 candidate mission currently in Phase A, with Launch in $\sim$2032. The aim of the mission is to complete a Gamma Ray Burst survey and monitor transient X-ray events. The University of Leicester is the PI institute for the Soft X-ray Instrument (SXI), and is responsible for both the optic and detector development. The SXI consists of two wide field, lobster eye X-ray modules. Each module consists of 64 Micro Pore Optics (MPO) in an 8 by 8 array and 8 CMOS detectors in each focal plane. The geometry of the MPOs comprises a square packed array of microscopic pores with a square cross-section, arranged over a spherical surface with a radius of curvature twice the focal length of the optic. Working in the photon energy range 0.3-5 keV, the optimum $L/d$ ratio (length of pore $L$ and pore width $d$) is upwards of 50 and is constant across the whole optic aperture for the SXI. The performance goal for the SXI modules is an angular resolution of 4.5 arcmin, localisation accuracy of $\sim$1 arcmin and employing an $L/d$ of 60. During the Phase A study, we are investigating methods to improve the current performance and consistency of the MPOs, in cooperation with the manufacturer Photonis France SAS. We present the optics design of the THESEUS SXI modules and the programme of work designed to improve the MPOs performance and the results from the study.
\end{abstract}


\section{INTRODUCTION}
\label{intro}
THESEUS, is a multi-instrument mission to detect and characterize Gamma-Ray Bursts (GRBs). It is an M5 candidate mission of the Cosmic Vision programme, consisting of three instruments, X/Gamma-ray Imaging Spectrometer (XGIS)\cite{xgisspie}, Soft X-ray Imager (SXI)\cite{pobspie} and the InfraRed Telescope (IRT)\cite{irtspie}.

There are two main scientific goals of the THESEUS mission. The first is to explore the Early Universe in particular the cosmic dawn and reionisation era, by unveiling a complete census of the Gamma-Ray Burst (GRB) population in the first billion years. The second is to perform an unprecedented deep monitoring of the X-ray transient Universe. THESEUS will specifically perform a study of global star formation up to z $\sim$10. In addition it will detect and study the primordial (pop III) star population and investigate the re-ionisation epoch, the interstellar medium (ISM) and the intergalactic medium (IGM), up to z $\sim$8 – 10. THESEUS will provide real-time triggers and accurate locations of GRBs (both long and short) and high-energy transients for follow-up with next-generation optical-NIR, radio, X-rays, TeV telescopes. Further details of the THESEUS science and instruments can be found in papers 11444-302\cite{laspie}, 303, 304 and 305 in these proceedings.

The SXI instrument on-board the THESEUS space craft comprises 2 identical modules, each being a wide-field, lobster eye X-ray telescope. The University of Leicester (UoL) are the PI institute for the SXI instrument on-board THESEUS, and is responsible for the optic and detector development and characterisation as well as leading the development of the software. Papers 11444-304 and 284 within these proceedings, provide further details of the different aspects of this instrument and their development. This paper details the development plan for the novel lobster eye optics in partnership with the company producing the Micro Pore Optics (MPOs), Photonis France SAS.

\section{Photonis France SAS}
\label{phot}
Photonis is a high-technology organization with over 75 years of experience in innovating, developing, manufacturing, and selling sensor technologies for detecting and amplifying very low light levels, charged particles or harsh radiation. Photonis products are used in a very wide range of applications from night vision imaging devices to medical and analytical instruments or in nuclear reactors. Since decades Photonis is at the forefront of Micro Channel Plate (MCP) detector technology and more recently, developing Micro Pore Optics for X-ray instruments on board of multiple space satellite missions. They produced all of the MPOs for the MIXS\cite{fras2} instrument on the ESA BepiColombo mission, the MXT\cite{gotz} instrument on the Chinese-French mission SVOM, the SXI instrument on the ESA-CAS mission SMILE\cite{smile} and for the LEXI\cite{lexi} instrument on NASA's moon lander payload.

The manufacturing of MPOs and MCPs is in many aspects very similar. However, MPOs for X-ray collimation or focussing differ from MCP technology in several ways. MCPs have round pores and are active charge amplifiers with $\sim$1000 V running through each one. MPOs on the other hand have square holes and are passive elements i.e. requiring no electronics. Standard square pores with widths of 10 $\mu$m, 20 $\mu$m, 40 $\mu$m, 83 $\mu$m or 720 $\mu$m and can be square packed and slumped like a lobster eye, e.g. SVOM's MXT and SMILE's SXI, or radially packed, e.g. BepiColombo's MIXS. 
 
\begin{figure}
	\centering
		\includegraphics[width=0.95\textwidth]{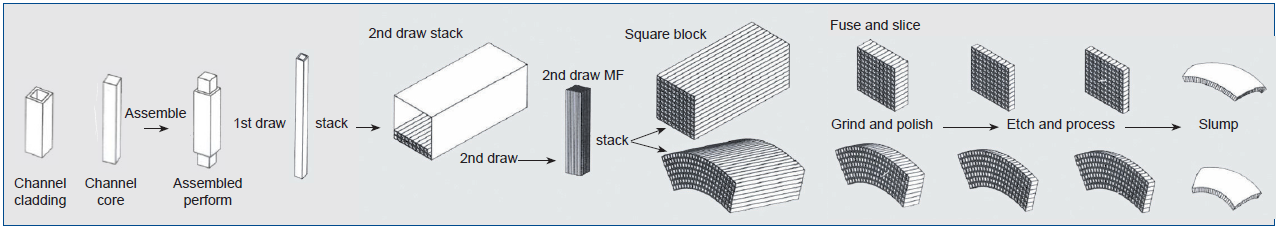}
			\caption{The MPO manufacturing process. (Credit: Photonis France SAS)}
	\label{photpro}
\end{figure}

MPOs are created in a multi stage production, usually about 40 main stages (Figure \ref{photpro}), but is summarised here. First a square core glass is surrounded by an inert cladding glass which will form the micro pore walls. This is then drawn in a tower to form a $\sim$1 mm wide, square fibre, which is then cut at a pre-fixed length and stacked to form a multifibre. The multifibre is then again drawn until it is $\sim$1 mm wide. Multifibres are then stacked together to form the basis of the MPO block that ultimately fits the outer dimensions of the required MPOs. The multifibre stack then is compressed under high temperature and pressure in a fusion process to make it into a solid block. Individual MPOs are then sliced from the block, followed by grinding and polishing down to the required thickness, defining the channel length. The MPOs are etched to remove the core glass and then slumped to a spherical or cylindrical shape. Finally they are iridium coated within the pores and a thin aluminium film is applied over the front pore apertures. The surfaces inside the vast amount of square channels operate as an assembly of ``micro-mirrors'' with a near-perfect flatness and a very low roughness.

\section{SXI design}
\label{sxi}
The lobster eye geometry for X-ray imaging was first introduced by Angel (1979)\cite{ang} and the use of tesselated slumped Micro Channel Plates (MCPs) in a lobster eye X-ray telescope has been pursued by several authors\cite{theory, wilks, fraspie, kaa}. This geometry can  provide a very large Field of View (FoV) and in addition, due to the low density of the glass used in the optics, has very low mass. A wide-field lobster eye optic configuration maintains the same unvignetted effective area across the entire FOV. This technology and geometry is currently being explored for multiple X-ray telescope missions including SVOM\cite{gotz} and SMILE\cite{smile}and future proposed missions such as TAP\cite{tap} and THESEUS\cite{thes}. For further details about lobster eye geometry and the distinctive point spread function produced by single MPOs and arrays of these devices, see paper 11444-159\cite{svomspie} in these proceedings.

\begin{figure}
	\centering
		\includegraphics[width=0.25\textwidth]{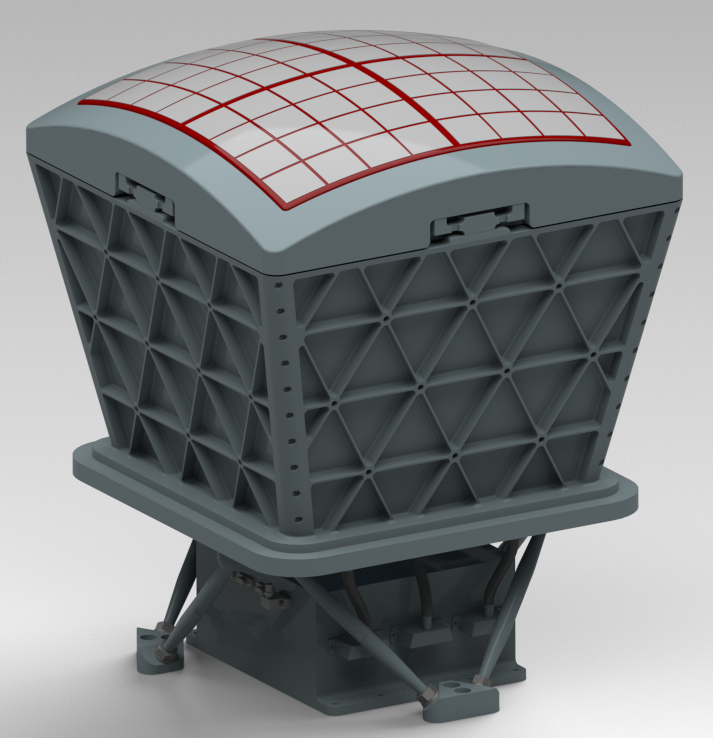}
		\includegraphics[width=0.25\textwidth]{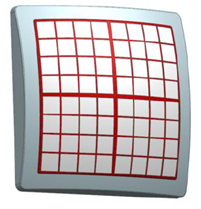}
			\caption{A CAD rendering of a THESEUS SXI module on the left and just the optic on the right.}
	\label{thesopt}
\end{figure}

The science goals of THESEUS make a wide-field lobster eye X-ray telescope a perfect candidate for the SXI instrument. The optics aperture for the SXI modules is formed by an array of 8 by 8 square, MPOs mounted on an aluminium spherical frame with a radius of curvature of 600 mm. Each module provides a FoV of 31$^{o}$ by 31$^{o}$ and combined a FoV of $\sim$0.25 steradians. For X-ray applications, a wide field lobster eye telescope working in the photon energy range 0.2-10 keV, has a constant and optimum $L/d$ ratio (length of pore $L$ and pore width $d$) of $\sim$50. For the optimum performance of the SXI at 1 keV, the $L/d$ needs to be 60, and thus for MPOs with a pore width of 40 $\mu$m the required thickness is 2.4 mm. Each MPO is 40 mm by 40 mm square with a 2 mm gap in between for bonding and a 1 mm overlap with the frame. This means that each MPO aperture is 38 mm by 38 mm. The frame is 420 mm square, which includes the MPOs and the surrounding supportive structure. The MPOs are mounted directly onto the frame using a silicon based adhesive. A CAD rendering of a single THESEUS module and a single optic are shown in Figure \ref{thesopt}. Behind the optics are a set of magnets forming an electron diverter, along with the heaters to maintain the optics at a constant temperature of 20$^{o}$C $\pm$5$^{o}$C during flight. The full optics assembly of each module will weigh $\sim$8.4 kg, and each module will weigh less than 35.9 kg. The 2 modules will be aligned on the sky parallel to the spacecraft Sun shade, with a 1$^{o}$ wide overlap, co-aligned with the IRT FoV to provide redundancy. Each SXI module focal plane comprises of 8 CMOS detectors in a 2 by 4 arrangement which are slightly tilted in order to simulate the required curved focal plane required for a lobster eye optic and to avoid vignetting, and to optimise the FoV imaging area. Further details of the SXI focal plane can be found in paper 11444-284 within these proceedings.

\section{Facilities at the University of Leicester}
\label{fac}
There are two X-ray optics testing facilities within UoL, the TTF and the VTF.

The Tunnel Test Facility (TTF) is a 28 m long X-ray beam test facility with both soft and hard X-ray sources and two imaging detectors, an MCP and a CCD. At the detector end of the facility an $\sim$2 x 1.5 m diameter chamber contains a system for positioning and rotating the optic under test and the detector systems which are independently controllable in 3 linear dimensions by in vacuum motors. Up to 6 individual MPOs on a rotating carousel can be tested in a single pump-down cycle. The detector end of the facility opens in to an ISO 5 clean room with dry cabinet storage for the MPOs and optic assemblies. The facility is capable of doing temperature dependent tests and can actively control the temperature of the optics under test.

\begin{figure}
	\centering
		\includegraphics[width=0.36\textwidth]{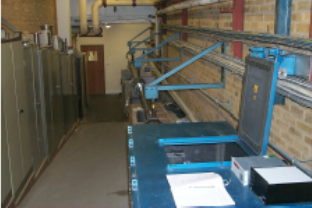}
		\includegraphics[width=0.3\textwidth]{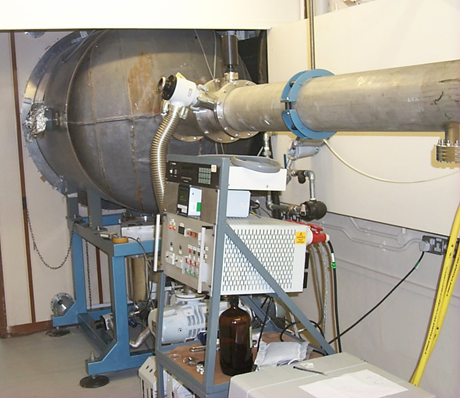}
			\caption{The TTF beam line at the University of Leicester. The source end is on the left and the detector chamber where the optics are mounted on the right.}
	\label{ttfpic}
\end{figure}

There are two X-ray sources which can be mounted within the source chamber. The soft X-ray source has recently been upgraded to enable a wider range of energies to be generated, from C-K (0.28 keV) to Ti-K (4.51 keV), covering the SXI bandpass. The hard X-ray source is capable of producing X-rays up to $\sim$70 keV. Either a Scandium or Tungsten anode can be used and can be configured to stimulate X-ray florescence of external targets providing a wide range of possible X-ray energies. Figure \ref{ttfpic} shows the source and the detector ends of the facility.

\begin{figure}
	\centering
		\includegraphics[width=0.35\textwidth]{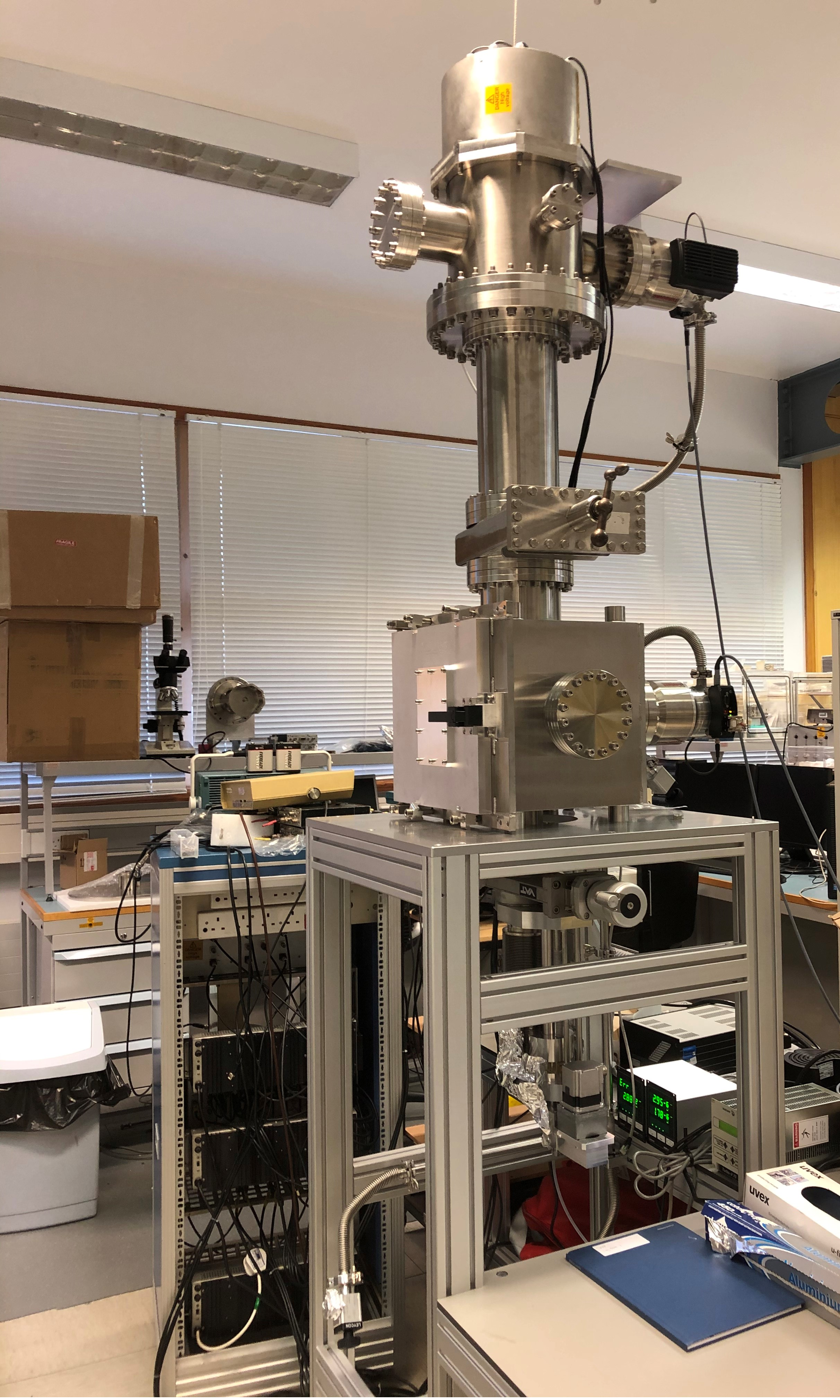}
			\caption{The VTF testing facility at the University of Leicester.}
	\label{vtfpic}
\end{figure}

The Vertical Test facility (VTF) is a 1.5 m tall X-ray beam facility with an electron gun to generate the X-rays of characteristic wavelengths at the bottom, a sample stage in the central chamber and an imaging MCP detector in the upper chamber. The facility has a vertical orientation so that the MPO optic is retained in place by gravity only eliminating mounting distortion and stress. In addition, MPOs with Al film on the convex surface can be tested without causing damage to the film. The MPO is tested in out-of-focus mode with the rear surface towards the source. A parallel beam from the X-ray source is created when the MPO is at (or close to) the expected focal point for a point source at infinity. By using a mask directly above the MPO with a specific sequence of holes, it is possible to select only the single reflection, double reflection and straight through rays which constitute the point-spread function of a lobster optic. From a single image of this type, it is possible to calculate the efficiency, radius of curvature, pointing direction of the pores and PSF of the MPO. The facility allows up to two MPOs to be tested a day giving a very fast turn around time and short feedback loops. This facility will be complete before the end of 2020 and an image of it in its current state is shown in Figure \ref{vtfpic}.

\section{Development plan}
\label{devp}
A Technology Development Activity (TDA) grant from ESA has been provided to improve the consistency of the MPOs produced by Photonis and to guarantee the science performance. The goal is to achieve a PSF FWHM for each module of 4.5 arcmin, though the science can still be achieved as long as the FWHM is below 7 arcmin. In order to achieve a module PSF FWHM of $<$7 arcmin, each individual MPO requires a PSF FWHM of $<$5 arcmin.

Photonis have produced MPOs with a FWHM of $\sim$6.5 arcmin in the past, but the manufacturing yield of high angular resolution MPOs in each production batch is too low for the number required for THESEUS and hence costly. The current average FWHM performance of standard Photonis MPOs is larger than that required for the THESEUS mission. In addition, there is a larger than acceptable variation in the radii of curvatures between the MPOs coming from a single batch, which will increase the size of the combined focus produced by an array of MPOs, as used in the SXI. The development activity is aiming to push the precision limits of the MPO manufacturing processes to produce large batches of MPOs with consistently higher angular resolution and the smallest possible variations in focal length.

Currently, only high precision mechanical metrology is applied during the MPOs production stages, and so, as purchased, the MPO performance is not known until they are X-ray tested. Using an MPO block specifically produced for this study and introducing X-ray and optical performance metrology at key production stages, we are confident that the block manufacturing and spherical slumping technique can be optimised and controlled to routinely produce MPOs which will reach the THESEUS performance goals.

The TDA is aiming to improve the MPOs from the very start of production. For this reason, Photonis have updated the fibre drawing facilities to the very latest state-of-the-art real-time automation, to monitor the fibres as they are drawn and thus remove human errors. This means that the fibre parameters, e.g. wall thickness, pore width, etc. are computer controlled, providing the most consistent production possible currently achievable. In addition, the fibre selection and integration steps have also been fully automated to ultimately provide a higher quality MPO. Because the production of the block has changed, to verify the quality improvement, initially flat MPOs from the new block will be tested at UoL before being returned to Photonis for slumping and coating (at a later stage).

In total 14 MPOs will be used within this TDA, 13 from the new block and a bare glass, 1.2 mm, flat MPO that was produced before the upgrade of the drawing facilities. This will serve as a reference-MPO for comparison between the old and new facilities. The 13 MPOs will be sliced from the new block, ground and polished down to 2.4 mm thickness and then etched. One of the 13 MPOs will then be iridium coated, and remain flat to compare the improvement of reflectivity between bare glass and coated. The 12 bare flat MPOs from the new block and the two reference MPOs will then be sent to UoL and tested under exactly the same conditions in the VTF to provide statistics and precise details on the efficiencies, fibre alignment, etc. with direct comparison between the old and new drawing techniques. It is expected that the fibre alignment will be much improved by the upgraded drawing facilities and thus one expects that the ultimate performance of the MPOs will be improved, quantified by the VTF measurements. 

The remaining twelve flat MPOs will be sent back to Photonis, of which six will be carefully slumped using the standard in house technique. After slumping, the six MPOs will be sent back to UoL for testing in both the VTF and TTF to measure the change in performance post slumping. Three of these slumped bare MPOs will be selected and subjected to no-contact metrology to characterise the slump profile. The six MPOs will be iridium coated at Photonis and X-ray tested again within the facilities at UoL to analyse the influence on the performance due to the coating, and compared to the flat coated reference MPO.

Based on these results, the slumping technique will be discussed with Photonis and possible processing changes will be considered for the second set of six MPOs. These will then be tested as the previous six for a direct comparison of the slumping processes. VTF, TTF and metrology data taken from the two different batches of six slumped MPOs will be compared and analysed in depth.

In addition to all of the above described X-ray testing and metrology, a few MPOs will be placed on a representative frame and then subjected to vibration and thermal tests. This to verify there has been no degradation (or even a possible improvement) in the MPOs environmental performance to ensure they are still fit for launch conditions.

\subsection{Current status}
\label{current}
Unfortunately, due to the Covid-19 pandemic, this work has been severely delayed. However, the new block has been drawn and fused at Photonis and is waiting to be sliced in to the individual 2.4 mm MPOs. It is hoped that the first flat MPOs will be received for testing at UoL by the end of March 2021 and the first slumped MPOs will be received in the summer of 2021. It is planned that this project will be completed by the end of 2021.

\section{SUMMARY AND CONCLUSION}
\label{sumcon}
The Transient High Energy Sources and Early Universe Surveyor (THESEUS) is an ESA M5 candidate mission aimed at studying GRBs and transient sources and completing an unprecedented survey of high z sources. Currently in Phase A, with Launch in $\sim$2032, it consists of three instruments. The University of Leicester is the PI institute for the SXI on-board THESEUS, leading the developments and characterisation in both the optics and detectors. Through the ESA funded TDA, the consistency of the Photonis MPOs is being improved in
order to meet the science and performance goals of the THESEUS mission. This development plan is a collaboration between the University of Leicester and Photonis France SAS.

\section{ACKNOWLEDGEMENTS}
\label{ack}
We acknowledge contributions from the SXI consortium and the wider THESEUS team to the development of the SXI. We also acknowledge funding from ESA (4000129734/20/NL/IB/ig) and several ESA member states, including from the UK Space Agency.

\bibliography{bobs}{}
\bibliographystyle{spiebib}
\end{document}